\documentclass[journal,onecolumn]{IEEEtran}
\IEEEoverridecommandlockouts
\usepackage{setspace} 
\usepackage[justification=centering]{caption}
\usepackage{cite}
\usepackage{amsmath,amssymb,amsfonts,amsthm,verbatim}
\usepackage{algorithmic}
\usepackage{graphicx}
\usepackage{textcomp}
\usepackage{xcolor}

\usepackage[utf8]{inputenc}
\usepackage[T1]{fontenc}

\DeclareMathOperator{\rank}{rk}
\DeclareMathOperator{\supp}{supp}

\DeclareMathOperator{\wt}{wt}
\DeclareMathOperator{\rows}{Rows}
\DeclareMathOperator{\prob}{P}
\DeclareMathOperator{\pir}{PIR}

\newtheorem{definition}{Definition}

\newtheorem{theorem}{Theorem}

\newtheorem{corollary}{Corollary}

\newtheorem{remark}{Remark}

\newif\ifcomment
\commenttrue
\commentfalse 

\begin{document}

\title{Code-Based Single-Server Private Information Retrieval: Circumventing the Sub-Query Attack}

\author{\IEEEauthorblockN{Neehar Verma and Camilla Hollanti}\\\thanks{This work was supported by the Academy of Finland (grant \#336005) and by the MSCA Doctoral Networks 2021, HORIZON-MSCA-2021-DN-01 (grant \#101072316).}
\IEEEauthorblockA{\textit{Department of Mathematics and Systems Analysis} \\
\textit{Aalto University, School of Science}\\
Espoo, Finland \\
\{neehar.verma, camilla.hollanti\}@aalto.fi}
}
\maketitle

\begin{abstract}
Private information retrieval from a single server is considered, utilizing random linear codes. Presented is a modified version of the first code-based single-server computational PIR scheme proposed by Holzbaur, Hollanti, and Wachter-Zeh in [Holzbaur \emph{et al.}, ``Computational Code-Based Single-Server Private Information Retrieval'', \emph{2020 IEEE ISIT}]. The original scheme was broken in [Bordage \emph{et al.}, ``On the privacy of a code-based single-server computational PIR scheme'', \emph{Cryptogr. Comm.}, 2021] by an attack arising from highly probable rank differences in sub-matrices of the user's query. Here, this attack is now circumvented by ensuring that the sub-matrices have negligible rank difference. Furthermore, the rank difference cannot be attributed to the desired file index, thereby ensuring the privacy of the scheme. In the case of retrieving multiple files, the rate of the modified  scheme is largely unaffected and at par with the original scheme. 
\end{abstract}

\section{Introduction}

Private information retrieval was first introduced by Chor \textit{et al.} in \cite{chor1995private, PIR} with the aim of enabling users to access data from a database or more generally from a distributed storage system while concealing the identity of the requested information from potentially untrusted servers. A trivial way to guarantee information-theoretic privacy is to download the entire database. Modern data storage systems may often contain a large number of big files and the trivial solution is infeasible in practice. 
More practical solutions which attempt to incur minimal communication overhead and related  capacity results for \emph{information-theoretically secure} PIR schemes are presented in  \cite{sun2017repcap,sun2017capacity, freij2017private, sun2018capacity, banawan2018capacity, tian2019capacity,holzbaur2022tit}. To enable information-theoretic privacy these works assume that the distributed storage system consists of sufficiently large subset of non-colluding servers. In practice, it may be difficult to decide for an appropriate level of collusion protection, and the more one protects the more penalty there is in terms of the achievable PIR rates. Moreover, if too many servers collude, user's privacy might be lost. For this reason, considering a single server or, equivalently, full collusion, becomes interesting.
In this case,  information-theoretic privacy can only be achieved by downloading all the files. As a more practical alternative, schemes which are \emph{computationally secure} have been examined in several works. Certain schemes, \emph{e.g.}, \cite{kushilevitz1997replication, lipmaa2005oblivious, gentry2005single},
make use of computationally hard problems in the realm of classical computers, such as the quadratic residuosity problem. Such schemes will be rendered insecure when quantum computing matures, since the underlying hard problems can be efficiently solved using quantum algorithms.

\subsection{Related work and contributions} 

In the realm of post-quantum security, lattice-based cryptography has emerged as a promising avenue, with \cite{aguilar2007lattice} proposing an efficient lattice-based computational PIR scheme. While this approach initially seemed robust, a practical vulnerability was unveiled in \cite{liu2016cryptanalysis}, specifically targeting databases with a limited number of elements. However, such a limitation may not pose a significant threat, given the prevalent use of databases with a large number of elements.

The introduction of the first fully homomorphic encryption (FHE) scheme in \cite{gentry2009fully} marked a breakthrough in cryptography. FHE was subsequently leveraged to construct a general PIR scheme in \cite{yi2012single}. Several other PIR schemes based on FHE are presented in \cite{kiayias2015optimal,aguilar2016xpir,lipmaa2017simpler,gentry2019compressible}. Schemes based on FHE offer computationally secure PIR, but may often come at the cost of a high communication complexity.

The scheme proposed in \cite{holzbaur2020isit} introduced the first code-based PIR scheme. We will shortly refer to this as the \emph{HHW~scheme}. In the HHW scheme, the server is queried using a matrix comprising corrupted codewords selected from a random linear code. The confidentiality of the desired file index is maintained through specifically crafted errors embedded in the query matrix, the decoding of which is known to be an NP-hard problem  \cite{Berlekamp1978}. Upon receiving the server's response, decoding exposes the errors, and projection onto a relevant vector subspace unveils the desired file. As these errors were initially picked by the user, they simply need to do erasure decoding. Notably, the HHW scheme, with carefully chosen parameters, achieves PIR rates comparable to the computational PIR schemes presented in \cite{aguilar2007lattice,yi2012single}.

For the proposed parameters in \cite[Section III.4]{aguilar2007lattice} the computational complexity is seen to be the complexity of matrix multiplications over the field $\mathbb{F}_{2^{60} + 325}$. For the HHW scheme with parameters achieving similar retrieval rates, the computational complexity is approximately equal to multiplication of matrices of similar size over a significantly smaller field $\mathbb{F}_{2^{29}}$. Another appealing feature of the HHW scheme lies in its ability to perform calculations over binary extension fields. Despite its merits, the security of the HHW scheme was called into question in \cite{bordage2020privacy}. The identified vulnerability enables an attacker to discern the secret by observing rank differences in sub-matrices of the query.

In \cite{Alfarano2023survey} the authors develop a code-based framework which formalizes several single-server PIR schemes. In this framework it is seen that any PIR scheme similar to the HHW scheme is susceptible to the sub-query attack. The authors in \cite{bodur2023ring} circumvent this attack by using non-free codes over rings. These non-free codes are constructed by applying the Chinese remainder theorem to codes which are non-Hensel lifts  \cite[Section IV, Corollary 7]{bodur2023ring}. 
As a main contribution of the present paper, we modify the HHW scheme to provide a mend against the sub-query attack and consequently to any similarly constructed scheme which is susceptible to this form of an attack.

When used to retrieve multiple files the rate of the modified scheme is largely unaffected and at par with the original HHW scheme. Hence, it preserves all the merits of the HHW scheme while now also ensuring privacy. 

\subsection{Notation}

We denote a finite field of size $q$  by $\mathbb{F}_q$ and $\mathbb{F}_q^\times=\mathbb{F}_q \setminus \lbrace 0 \rbrace$. The extension field $\mathbb{F}_{q^s}$ can be seen as a vector space of dimension $s$ over $\mathbb{F}_q$. For a set of linearly independent vectors $\Gamma = \lbrace \gamma_1, \dots , \gamma_v\rbrace \subset \mathbb{F}_{q^s}$ we denote by $\langle \gamma_1, \dots , \gamma_v \rangle_{\mathbb{F}_q} \subset \mathbb{F}_{q^s}$ the vector subspace of dimension $v$ over $\mathbb{F}_q$. Define the corresponding projection map, $\psi_\Gamma : \mathbb{F}_{q^s} \to \langle \gamma_1, \dots , \gamma_v \rangle_{\mathbb{F}_q}$.

For a vector $x \in \mathbb{F}_q^t$ and an ordered set $J \subset [m]=\{1,\ldots,m\}$ of size $t$ we define $\phi_J: \mathbb{F}_q^t \to \mathbb{F}_q^m$  to be the extension of $x$ with zeroes at indices $j \not\in J$. \emph{E.g.,} for $J=\lbrace 1,3\rbrace$ and $m = 5$, $\phi_J([x_1,x_2]) = [x_1, 0, x_2,0,0]$. For a set $I \subseteq [n]$ we denote the complement of this set by $\bar I = [n] \setminus I$.

\section{Outline of the HHW PIR scheme}

In this section, we summarize the setup of the original scheme in \cite{holzbaur2020isit}.

\subsection{System Model}
We are concerned with a single-server data storage containing $m$ files of size $L \times \delta$ over $\mathbb{F}_{q}$, where $\delta = (n-k)(s-v)$ and the parameters $n,k,s,v$ are as specified below. The data content on this server is denoted by $X \in \mathbb{F}_q^{L \times m\delta}$.

\subsection{Queries}
To construct the queries, the user samples uniformly at random the following:
\begin{itemize}
    \item A random linear code $C \subset \mathbb{F}_{q^s}^n$ of dimension $k$ and an information set $I$ of $C$.
    \item A matrix $D \in \mathbb{F}_{q^s}^{m\delta \times n}$ such that each row of $D$ is a codeword in $C$.
    \item A basis $\Gamma = \lbrace \gamma_1, \dots , \gamma_s\rbrace$ of $\mathbb{F}_{q^s}$ over $\mathbb{F}_{q}$, and the vector subspaces $V = \langle \gamma_1, \dots , \gamma_v \rangle_{\mathbb{F}_q}$ and $W = \langle \gamma_{v+1}, \dots , \gamma_s \rangle_{\mathbb{F}_q}$. 
\end{itemize}
We construct the matrix $E = \phi_{\bar I}(E_0)\in V^{m\delta \times n}$ where $E_0 \in V^{m\delta \times (n-k)}$, and the full-rank matrix $\Delta = \phi_{\bar I}(\Delta_0)\in W^{\delta \times n}$ where $\Delta_0 \in W^{\delta \times (n-k)}$. 
Finally, the user constructs the query $$Q^i = D + E + \Delta \otimes e^i$$

\begin{figure}[htbp]
\centering
\includegraphics[width=100mm,keepaspectratio]{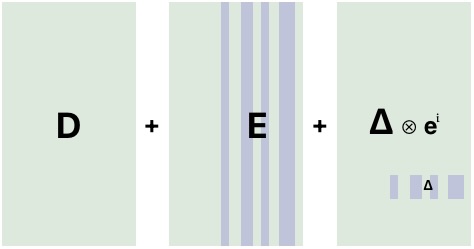}
\caption{Illustration of the query matrix $Q^{i}$.}
\label{fig1}
\end{figure}

\subsection{Retrieval}
Decompose $Q^i$ as the stack of of sub-matrices $Q^i_1, \dots Q^i_m \in \mathbb{F}_{q^s}^{\delta \times n}$. The server upon receiving the query responds with
\begin{align*}
    A^i & = X\cdot Q^i \\
    & = [X^1|\cdots |X^m] \cdot [Q^i_1|Q^i_2|\cdots |Q^i_m]^T \\
    & = \sum_{j=1}^m X^j\cdot Q^i_j \\
    & = \sum_{j=1}^m X^j\cdot D_j + \sum_{j=1}^m X^j\cdot E_j  + X^i\cdot \Delta.
\end{align*}
The rows of the matrix $\sum_{j=1}^n X^j\cdot D_j$ lie in $C$ and the rows of $\sum_{j=1}^n X^j\cdot E_j  + X^i\cdot \Delta$ have support $\bar I$ . Therefore by erasure decoding we can obtain $B^i = \sum_{j=1}^n X^j\cdot E_j  + X^i\cdot \Delta$.
 We can then project onto the space W and get $\psi_W(B^i) = X^i\cdot \Delta$. Since $\Delta$ has full rank we can recover the desired file $X^i$.
 
\subsection{Analysis of the scheme} Let us next recall the rate achievable by the HHW scheme. 

\begin{definition} The \emph{PIR rate} of a given scheme is (informally) defined as 
$$R_{\pir}=\frac{\mathrm{The\ size\ of\ the\ desired\ file(s)}}{\mathrm{The\ total\ download\ size}}.$$
\end{definition}

\begin{theorem}\cite[Thm~1]{holzbaur2020isit} The rate of the HHW scheme is 
$$R_{\pir} = \frac{L\delta \log(q)}{(m\delta n + Ln) \log(q^s)}=\frac{L\delta }{ns(m\delta  + L)}.$$
\end{theorem}

\begin{corollary}\cite[Cor.~1]{holzbaur2020isit} Assume $L>>\delta m$, \emph{i.e.}, the size of the files is large compared to the number of them and we can safely ignore the upload cost. Then the rate of the scheme is 
\begin{align*}
    R_{\pir} & \approx \frac{\delta}{ns}
     = 1 - \frac{k + \frac{v}{s}(n-k)}{n}.
\end{align*}

\end{corollary}

\section{Outline of the Sub-Query Attack}

Consider the sub-matrices $Q^i[j]$ of the received query where the rows $[(j-1)\delta +1,j\delta]$ of $Q^i$ are deleted, $j \in [m]$.

It is shown in \cite{bordage2020privacy} that we can decompose $$\mathbb{F}_{q^s}^n = C ~ \oplus ~ \phi_{\bar I}(V^{n-k}) ~ \oplus  ~ \phi_{\bar I}(W^{n-k}).$$
Due to this fact the $\mathbb{F}_q$-rank of a sub-matrix $$\rank(Q^i[j]) = \rank(D[j]+E[j]) + \rank(\Delta \otimes e^i[j]).$$ 
For $j \neq i$, $\rank(Q^i[j]) = \rank(D[j]+E[j]) + \delta$, and $\rank(Q^i[i]) = \rank(D[i]+E[i]) \leq ns - \delta$.

The attack involves computing the $\mathbb{F}_q$-rank of all $m$ sub-matrices $Q^i[j]$ and discerning the desired file index due to the low rank of $Q^i[i]$. Discerning the desired file index is only possible if $\rank(Q^i[i]) < \rank(Q^i[j])$ for all $j \neq i$, that is,  the attack fails if $\rank(D[j] + E[j]) < ns - 2\delta$. In \cite{bordage2020privacy} the authors prove that the probability 
\begin{align*}
    p ~ & := \prob(\rank(D[j] + E[j]) < ns - 2\delta) \\
    & \leq \left[ ns - \delta \atop ns - 2\delta \right]_q q^{-\delta^2(m-1)} 
     \leq q^{(\delta + 1)(ns - 2\delta) - \delta^2(m-1)}.
\end{align*}
As long as $(\delta + 1)(ns - 2\delta) < \delta^2(m-1)$ this probability is meaningful. 
In other words, when $m > 1 + \frac{(\delta + 1)(ns - 2\delta)}{\delta^2}$ the attack can discern the desired file index with high probability, thereby breaking the scheme for an unbounded number of files.

\section{Modified HHW Scheme}
In the original scheme \cite{holzbaur2020isit} the secret in the query came from the standard unit vector $e^i$. The attack in \cite{bordage2020privacy} with high probability can reveal this secret due to the fact that the standard unit vector has low weight. In the modified scheme each query is constructed with a secret of high weight. As a result each query itself does not allow us to retrieve a file, but a specifically constructed set of queries allows us to retrieve multiple desired files privately. 


The scheme is constructed over the same field and with the same parameters similarly as  in \cite{holzbaur2020isit}. Retrieval of the files is also done in the same manner as the original scheme. In the modified version of the scheme the user wishes to retrieve $f$ files, represented by $J = \lbrace j_1, \dots, j_f \rbrace \subset [m]$.

\subsection{Queries}
\begin{figure}[htbp]
\centering
\includegraphics[width=100mm,keepaspectratio]{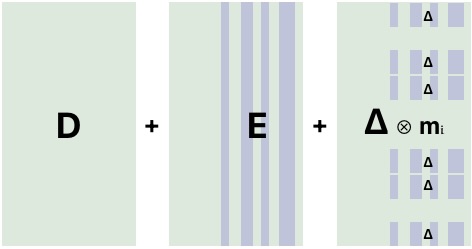}
\caption{Illustration of the query matrix $Q^{m_i}$.}
\label{fig2}
\end{figure}

The user samples uniformly at random a full-rank matrix $\Tilde{M} \in (\mathbb{F}_q^\times)^{f\times (f+1)}$ and a full-weight vector $\beta \in (\mathbb{F}_q^\times)^m$. 
The user then constructs $$M = \begin{bmatrix}
    \Tilde{M} \\
    0 ~ 0 \cdots 0 ~ 1
\end{bmatrix} \begin{bmatrix}
    e^{j_1} \\
    \vdots \\
    e^{j_f} \\
    \beta
\end{bmatrix} = \begin{bmatrix}
    m_{1} \\
    \vdots \\
    m_{f} \\
    \beta
\end{bmatrix}.$$
\begin{remark}
    Note that $m_i$'s are constructed such that $wt(m_i)\geq m-f$. Optimally, $m_i$'s can be constructed such that they have full weight, $wt(m_i) = m$.
\end{remark}
For each individual query $Q^{m_i}$ the user  samples independently and uniformly at random:
\begin{itemize}
    \item A random linear code $C \subset \mathbb{F}_{q^s}^n$ of dimension $k$ and an information set $I$ of $C$.
    \item A matrix $D \in \mathbb{F}_{q^s}^{m\delta \times n}$ such that each row of $D$ is a codeword in $C$.
    \item A basis $\Gamma = \lbrace \gamma_1, \dots , \gamma_s\rbrace$ of $\mathbb{F}_{q^s}$ over $\mathbb{F}_{q}$, and the vector subspaces $V = \langle \gamma_1, \dots , \gamma_v \rangle_{\mathbb{F}_q}$ and $W = \langle \gamma_{v+1}, \dots , \gamma_s \rangle_{\mathbb{F}_q}$. 
\end{itemize}
We construct the matrix $E = \phi_{\bar I}(E_0)\in V^{m\delta \times n}$ where $E_0 \in V^{m\delta \times (n-k)}$ and the full-rank matrix $\Delta = \phi_{\bar I}(\Delta_0)\in W^{\delta \times n}$, where $\Delta_0 \in W^{\delta \times (n-k)}$. Finally, the user constructs the queries (see Fig. \ref{fig2}) $$Q^{m_i} = D + E + \Delta \otimes m_i,\ 
Q^\beta = D + E + \Delta \otimes \beta.$$

\subsection{Retrieval}
Decompose each $Q^{m_i}$ as the stack of the sub-matrices $Q^{m_i}_1, \dots, Q^{m_i}_m \in \mathbb{F}_{q^s}^{\delta \times n}$. The server upon receiving the query responds with
\begin{align*}
    A^i & = X\cdot Q^{m_i} \\
    & = [X^1|\cdots |X^m] \cdot [Q^{m_i}_1|\cdots |Q^{m_i}_m]^T 
     = \sum_{j=1}^m X^j\cdot Q^{m_i}_j \\
    & = \sum_{j=1}^m X^j\cdot D_j + \sum_{j=1}^m X^j\cdot E_j  +  X \cdot \Delta \otimes m_i.
\end{align*}
The rows of the matrix $\sum_{j=1}^n X^j\cdot D_j$ lie in $C$ and the rows of $\sum_{j=1}^n X^j\cdot E_j  + X \cdot (\Delta \otimes m_i)$ have support $\bar I$ . Therefore by erasure decoding we can obtain $B^{m_i} = \sum_{j=1}^n X^j\cdot E_j  + X \cdot (\Delta \otimes m_i)$. We can then project onto the space W and get $\psi_W(B^{m_i}) = X \cdot (\Delta \otimes m_i)$. Since $\Delta$ has full rank we can recover $ X \cdot (I_{\delta \times \delta} \otimes m_i) $. Compiling the retrieved information by stacking the decoded response matrices we obtain $X \cdot (I_{\delta \times \delta} \otimes M)$. Since $\Tilde{M}$ has full rank, we can obtain  
$$X \cdot \left(I_{\delta \times \delta} \otimes \begin{bmatrix}
    e^{j_1} \\
    \vdots \\
    e^{j_f}
\end{bmatrix} \right).$$
The $f$ blocks of this matrix give us our desired files $X^{j_1},\dots , X^{j_f}$.

\subsection{Analysis of the modified scheme}
Let us now look into the modified scheme in more detail. The size (in bits) of each file we desire is $L\delta \log_2(q)$. The size of each of the first queries $Q^{m_i}$ and $Q^\beta$ is $m\delta n \log_2(q^s)$ with response size $Ln\log_2(q^s)$.

\begin{theorem}
  The PIR rate of the scheme is $$R_{\pir} = \frac{f L\delta \log(q)}{(f+1)(m\delta n + Ln) \log(q^s)}.$$
\end{theorem}

\begin{corollary}\label{approxRate}
    Assume $L>>\delta m$, \emph{i.e.}, the size of the files is large compared to the number of them and we can safely ignore the upload cost. Then the rate of the scheme is,
    \begin{align*}
      R_{\pir} & \approx \frac{f\delta}{(f+1)ns}
       = \frac{f}{f+1}\left(1 - \frac{k + \frac{v}{s}(n-k)}{n}\right) .
    \end{align*}
\end{corollary}

For a large number of retrieved files $f$ the rate is approximately the same as in \cite{holzbaur2020isit}. 

\subsection{Example}
Suppose we want to privately retrieve a single file with index $i$ from the server with elements in $\mathbb{F}_2$.  We construct $$M = \begin{bmatrix}
    1 ~ 1 \\
    0 ~ 1
\end{bmatrix} \begin{bmatrix}
    e^{i} \\
    1 ~ 1 \cdots 1 ~ 1
\end{bmatrix} = \begin{bmatrix}
    1 \cdots 1 ~ 0 ~ 1 \cdots 1 \\
    1 \cdots 1 ~ 1 ~ 1 \cdots 1 
\end{bmatrix} = \begin{bmatrix}
    m_i \\
    \beta
\end{bmatrix},$$
 and send the queries
$$Q^{m_i} = D + E + \Delta \otimes m_i,\ 
Q^\beta =D + E + \Delta \otimes \beta.$$

From the server's response, we obtain 
$$\begin{bmatrix}
    I_{L \times L} ~ I_{L \times L}
\end{bmatrix} \begin{bmatrix}
    \sum_{j \neq i} X^j \\
    \sum_{j = 1}^m X^j 
\end{bmatrix} = X^i.$$
The achieved rate as approximated in Corollary \ref{approxRate} is $$R \approx \frac{\delta}{2ns}.$$
\begin{remark}
    The response from the query $Q^\beta$ if stored can be reused for subsequent private file retrievals, making the rate of the subsequent retrievals equivalent to the rate in \cite{holzbaur2020isit}.
\end{remark} 

\section{Thwarting the Borgade--Lavauzelle attack}

Let us now see in more detail how to circumvent the sub-query attack in   \cite{bordage2020privacy}.

\subsection{Modified scheme vs. original attack}
For the original queries from \cite{holzbaur2020isit},
$$Q^i = D + E + \Delta \otimes e^i,$$
it was shown in \cite{bordage2020privacy} that for a desired file $X^i$, the sub-matrices of the query have $\mathbb{F}_q$-rank $$\rank(Q^i[j]) = \rank(D[j]+E[j]) + \delta$$ for $j \neq i$ and $$\rank(Q^i[i]) = \rank(D[i]+E[i]) \leq ns - \delta.$$
These sub-matrices with high probability have a discernible rank difference, allowing the server to reveal the desired file index.

Consider the case of the modified scheme with queries  
$$Q^{m_i} = D + E + \Delta \otimes m_i.$$
 The sub-matrices of the query have $\mathbb{F}_q$-rank $$\rank(Q^i[j]) = \rank(D[j]+E[j]) + \delta$$ for all $j \in [m]$. Therefore, the server cannot ascertain the desired file index by computing the sub-matrix ranks.
\begin{remark} Since $D$, $E$, $\Delta$,  and the code $C$ and its information set are chosen independently and randomly for each query, the server cannot reconstitute the initial queries before the row operations performed by the matrix $\Tilde{M}$ in order to perform the sub-matrix rank attack.
\end{remark}

\subsection{Modified scheme vs. modified attack}
A natural way to extend the attack in \cite{bordage2020privacy} to the modified scheme could be to compute the $\mathbb{F}_q$-ranks of sub-matrices $Q^{m_i}[J]$, where $J \subset [m]$ and $|J| = wt(m_i)$. For all such $J$ we have
$$\rank(Q^{m_i}[J]) \leq (m-\wt(m_i))\delta. $$
Let $I = \supp(m_i)$. Then
$$\rank(Q^{m_i}[I]) = \rank(D[I]+E[I]) \leq ns - \delta.$$ Otherwise, 
 for $J \neq \supp(m_i)$,
$$\rank(Q^{m_i}[J]) = \rank(D[J]+E[J]) + \delta\leq ns.$$
The secret $m_i$ is only discernible by the attacker if  $\rank(D[J] + E[J])$ does not shrink too much with respect to that of $I$. 
If we construct $m_i$ such that $(m-\wt(m_i))\delta < ns - \delta$ then $\rank(Q^{m_i}[I])$ and $\rank(Q^{m_i}[J])$ are indistinguishable. 
That is, we want $m_i$ such that $$\wt(m_i) \geq m + 1 - \frac{f}{(f+1)R_{\pir}}.$$

\begin{remark}
    We can always construct $m_i$ that satisfy the above inequality. Optimally we can construct $m_i$'s such that $\wt(m_i)~=~m$. 
\end{remark}

In the \emph{avoidable} case where $\wt(m_i) < m + 1 - \frac{f}{(f+1)R_{\pir}}$ we can determine a bound on the probability that the attack is unsuccessful. The failure  probability is 
$$p = \prob(\rank(Q^{m_i}[J]) \leq ns - \delta)=\prob(\rank(D[J] + E[J]) \leq ns - 2\delta)$$
where for each query, the rows of $D+E$ are vectors chosen uniformly at random from $\mathcal{U} = C ~ \oplus ~ \phi_{\bar I}(V^{n-k}) $. Keeping notation consistent with \cite{bordage2020privacy}, we represent the set of the rows of $D[J] + E[J]$ (seen as vectors of length $ns$ over $\mathbb{F}_q$) by $\rows(D[J] + E[J])$.
The probability we want to compute is hence
$$
p = \prob(\exists \mathcal{A} \subset \mathcal U, \dim(\mathcal{A}) = ns - 2\delta ~|~  
\forall y\in \rows(D[J]+E[J]), y \in \mathcal{A}).
$$
By the union bound, we have
\begin{align*}
    p & \leq \sum_{\mathcal{A}\in Gr_{\mathcal{U}}(ns-2\delta)} \prob(\forall y\in \rows(D[J]+E[J]), y \in \mathcal{A}) \\
    & \leq \sum_{\mathcal{A}\in Gr_{\mathcal{U}}(ns-2\delta)} \prod_{t=1}^{(m - wt(m_i))\delta}\prob(y \in \mathcal{A} | y \leftarrow \mathcal{U}) \\
    & \leq \left[ ns - \delta \atop ns - 2\delta \right]_q q^{-\delta^2(m-wt(m_i))} \\
    & < q^{(\delta + 1)(ns - 2\delta) - \delta^2(m-wt(m_i))}
\end{align*}
where $Gr_{\mathcal{U}}(ns-2\delta)$ denotes the set of $(ns-2\delta)$-dimensional subspaces included in $\mathcal{U}$. This upper bound is meaningful when $(\delta + 1)(ns - 2\delta) \leq \delta^2(m-wt(m_i))$. To this end, we define $$m_0 := \wt(m_i) + \left\lceil \frac{(\delta +1)(ns-2\delta)}{\delta^2} \right\rceil.$$
Then for a database with number of files $m \geq m_0$ there exists an algorithm which can discern the secret $m_i$ from the query $Q^{m_i}$ with probability at least $1 - q^{-(m-m_0)\delta^2}.$ The running time for this algorithm is in $$\mathcal{O}\left(\left( m \atop \wt(m_i)\right)(m-\wt(m_i))(ns)^3\right)$$
since it involves finding the $\mathbb{F}_q$-rank of $\left( m \atop \wt(m_i)\right)$ submatrices of size $(m-\wt(m_i))\delta\times ns$.
\begin{remark}
    This running time is valid under the worst-case assumption that the attacker knows $\wt(m_i)$. In the case where $\wt(m_i)$ is unknown to the attacker, the running time for the algorithm is significantly higher.
\end{remark}

\section{User complexity}
The complexity of generating the queries in the modified scheme is inherently the same as in the HHW scheme barring the added complexity which comes from the Kronecker product $\Delta \otimes m_i$. 
The user therefore has to perform an additional $\wt(m_i)$ $\mathbb{F}_q$-scalar multiplications with $\Delta$ to generate a query. Since $m_i \in \mathbb{F}_q^{m}$ there is a pigeonholing of these scalar multiplications when $wt(m_i) > q$.

Therefore the number of unique computations required for generation of the query is in $\mathcal{O}(q\delta^2)$.
\begin{remark}
 To ensure security against attacks mentioned in \cite[Section V]{holzbaur2020isit} $q$ and $s$ must be sufficiently large. Keeping in mind the trade-off between user complexity and security a suitable choice for $q$ is $q=32$. Since $q$ and $\delta$ are magnitudes of order lesser than $m$, the additional user complexity for query generation (even when $\wt(m_i) \approx m$) is not too high.
\end{remark}

\section{Example and comparison}

\begin{figure}[htbp]
\centering
\includegraphics[width=180mm,keepaspectratio]{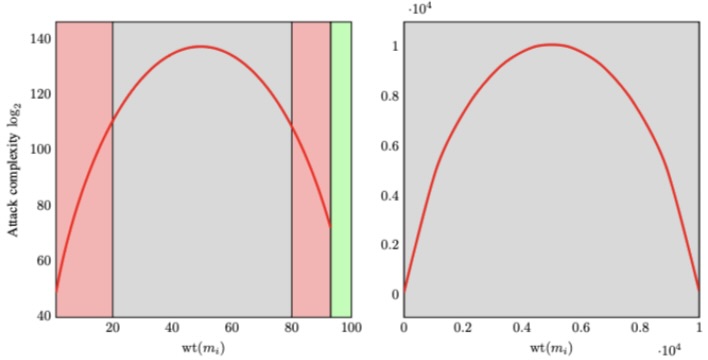}
\caption{Attack complexity vs. $\wt(m_i)$, $m=100$ and $m=10000$.}
\label{fig3}
\end{figure}

Consider the PIR scheme constructed with the parameters $$q=32, n=100, k=50, s=32, v=24, f = 1.$$
This gives us
$\delta = 400 \textit{ and } R_{\pir} \approx \frac{1}{16}.$ 
Consider  databases containing $m=100$ and $m=10000$ files respectively. The complexity of the modified attack with respect to the weight of the secret $m_i$ is illustrated in Figure \ref{fig3}.

In Figure \ref{fig3} the red regions highlight the domain in which the modified attack can with high probability discern the secret.

The red curve shows the number of computations required by the attack algorithm. 

The gray regions show the domain in which the the attack --- although possible --- is deemed infeasible (number of computations exceeding $2^{100}$). 

The green regions ($\wt(m_i) \geq m + 1 - \frac{f}{(f+1)R_{\pir}} = 93$ and  $9993$ respectively) highlight the region in which the attack is rendered impossible. To thwart the modified attack we can \emph{always} construct $m_i$ such that they lie in this desirable green region.  
\begin{remark}
    Regardless of the weight of $m_i$ the additional user complexity for query generation is in $\mathcal{O}(q\delta^2)$, allowing us to choose $m_i$ in the green region with negligible  impact on the user's complexity.
\end{remark}
The rates for other suitable, secure parameter choices as shown in \cite[Section VI]{holzbaur2020isit} are given in Table \ref{table1}.
\begin{table}
\centering
  \begin{tabular}{c c c c c c c|c}
    $q$ & $s$ & $v$ & $n$ & $k$ & $\delta$ & $f$ & $R_{\pir}$ \\
    \hline 
    16 & 32 & 31 & 100 & 50 & 50 & 1 & 1/128 \\
       & & & & & & 4 & 1/80 \\
       & & & & & & 32 & 1/66 \\
    \hline
    16 & 32 & 16 & 100 & 50 & 800 & 1 & 1/8 \\
       & & & & & & 4 & 1/5 \\
       & & & & & & 32 & 8/33 \\
    \hline
    32 & 32 & 31 & 100 & 50 & 50 & 1 & 1/128 \\
       & & & & & & 64 & 1/65 \\
    \hline
    32 & 32 & 26 & 100 & 50 & 300 & 1 & 3/64 \\
       & & & & & & 32 & 1/11 \\
    \hline
    32 & 32 & 24 & 100 & 50 & 400 & 1 & 1/16 \\
       & & & & & & 8 & 1/9 
\end{tabular}  
\caption{PIR rates as in Corollary \ref{approxRate}.}
\label{table1}
\end{table}

The ring-based PIR protocol in \cite{bodur2023ring} has rate
$$R_{\pir} \approx \frac{r}{2ns}\cdot \frac{\log(m')}{\log(m)}$$
where $r \leq s$.
Due to this limitation on $r$ the rate of this scheme cannot exceed $\frac{1}{2n}$. The scheme presented in this paper has no such limitation and can therefore outperform the ring-based scheme.

\section{Conclusions}

The modified scheme with suitably constructed secrets $m_i$ is able to circumvent the Borgade--Lavauzelle attack. The modified scheme preserves all the merits of the original HHW scheme, allowing for low computational complexity by performing calculations over binary extension fields. While this scheme is able to thwart the Borgade--Lavauzelle attack, there may be other computationally feasible attacks --- unknown thus far --- which could break it.  

The rate of this scheme coincides with that of the original HHW scheme when retrieving a large number of files. Improving the rate for a small number of files is an avenue which should be explored.

\section*{Acknowledgments}

The authors would like to thank \c{S}. Bodur, R. Freij-Hollanti, E. Mart\'inez-Moro, and D. Ruano for useful discussions.

\bibliographystyle{IEEEtran}
\bibliography{cPIR,IT_PIR}

\end{document}